# Cellular Network Traces Towards 5G: Usage, Analysis and Generation

Francesco Malandrino *Member, IEEE,* Carla-Fabiana Chiasserini *Senior Member, IEEE,*
Scott Kirkpatrick *Life Fellow, IEEE*

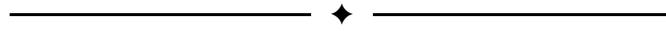

**Abstract**—Deployment and demand traces are a crucial tool to study today's LTE systems, as well as their evolution toward 5G. In this paper, we use a set of real-world, crowdsourced traces, coming from the WeFi and OpenSignal apps, to investigate how present-day networks are deployed, and the load they serve. Given this information, we present a way to generate *synthetic* deployment and demand profiles, retaining the same features of their real-world counterparts. We further discuss a methodology using traces (both real-world and synthetic) to assess (i) to which extent the current deployment is adequate to the current *and future* demand, and (ii) the effectiveness of the existing strategies to improve network capacity. Applying our methodology to real-world traces, we find that present-day LTE deployments consist of multiple, entangled, medium- to large-sized cells. Furthermore, although today's LTE networks are overprovisioned when compared to the present traffic demand, they will need substantial capacity improvements in order to face the load increase forecasted between now and 2020.

## 1 INTRODUCTION

Mobile devices, mobile cloud services and the upcoming Internet-of-things will foster an impressive growth in mobile data usage. As a consequence, mobile network operators (MNOs) need to take action to prevent their networks from being choked by data demand. Both the LTE-Advanced (LTE-A) specification [1] and the ongoing next-generation mobile networks design efforts [2] offer multiple options for mobile technology evolution. Prominent ones include multi-tier network densification (e.g., micro- and femtocells), the usage of multiple access technologies (e.g., Wi-Fi), coordinated transmissions, and the usage of new parts of the spectrum.

Such an abundance of potential solutions can, however, become a problem itself. Indeed, we need to establish (i) how the capacity of mobile networks compare to the traffic demand they have to serve and, more importantly, (ii) which *action* (if any), among the plethora of possible ones, mobile operators ought to take in order to cope with the expected growth in mobile data in the most effective (and profitable) way.

Real-world traces are a powerful tool to address these issues. Compared to traditional traces collected by individual mobile operators, *crowdsourced* traces, obtained with the assistance of individual users, are especially useful; indeed, their information is both more complete (as it includes multiple networks and network types, e.g., cellular and Wi-Fi) and more detailed (including the position of individual users and the app they use).

In this paper, we use two such crowdsourced traces, coming from the WeFi [3] and OpenSignal [4] apps and covering the San Francisco Bay Area, California – a heterogeneous zone encompassing dense urban, suburban and even rural zones. Using these traces, we endeavor to characterize present LTE networks and understand their future.

In addition to studying the real-world traces at our disposal, we also present a general methodology to generate new, *synthetic* traces exhibiting the same features. Synthetic traces are needed in two main cases. First and most obvious, when no real-world information is available, e.g., geographical areas served by operators that do not disclose any deployment or demand information. Furthermore, even when real-world traces are available, having multiple synthetic traces is useful to perform performance evaluation under a wider, more varied set of traffic loads.

There are four especially challenging tasks that we tackle in our study:

- cleaning, processing and combining the large-scale datasets we work with, complementing each other when they do not overlap and cross-checking them when they do, as well as verifying our findings against a third-party data source [5];
- deriving stochastic deployment and demand models that capture and reproduce the features of their real-world counterparts, both globally (e.g., yield a similar number of base stations) and locally (e.g., exhibit the same distance between base stations);
- integrating the information coming from the traces – be them real or synthetic – with external information like propagation models, experimental rate measurements and FCC license records, in order to obtain a comprehensive, reliable system representation;
- separately modeling each of the strategies foreseen to increase network throughput, so as to study their effect on the system performance.

The methodology we present in our paper is able to deal with all the tasks outlined above, and works unmodified with both real-world and synthetic traces. In particular, it

- *F. Malandrino and C.-F. Chiasserini are with Politecnico di Torino, Italy. C.-F. Chiasserini is also a Research Associate at IEIIT-CNR, Torino, Italy. S. Kirkpatrick is with the Hebrew University of Jerusalem, Israel.*

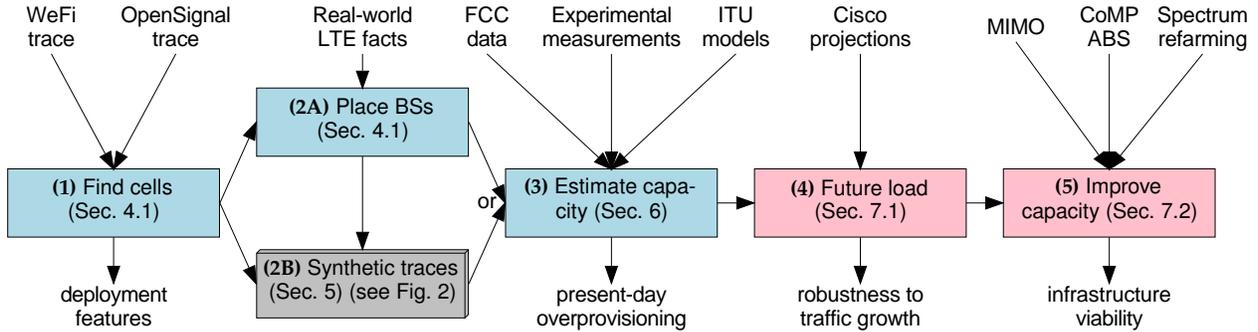

Fig. 1. The processing steps we perform (represented by boxes), the input data they use (arrows entering the boxes), and the information we obtain (arrows exiting from the boxes). Blue boxes deal with today's networks, while pink boxes refer to their future evolution. Steps 1–2A need to be performed on real-world traces; steps 3–5, on the other hand, can be performed either on real-world traces (as processed in step 2A) or on synthetic demand and deployment data, resulting from step 2B. Step 2B itself consists of multiple sub-steps, as detailed in Fig. 2.

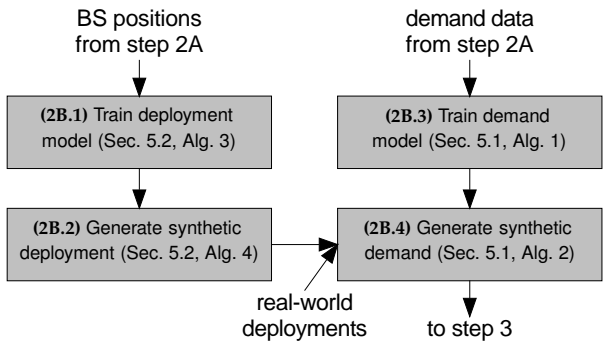

Fig. 2. Detail of step 2B in Fig. 1: how synthetic traces are generated. Models are trained with the information coming from step 2A in Fig. 1. Synthetic demands (step 2B.4) can be generated either using synthetic deployments, i.e., the output of step 2B.2, or from real-world deployment information. The output of step 2B.4, i.e., the synthetic trace, can then be processed as in step 3 of Fig. 1.

generates a network deployment that accurately matches the one reported by the third-party project in [5].

The main steps of our methodology are summarized in Fig. 1, which also describes the input and output of each individual step. In steps 1–2A we extract demand and deployment information from real-world traces, such as the WeFi and OpenSignal traces we work with. This information can be used, as shown in step 2B (further detailed in Fig. 2), to create synthetic traces exhibiting the same features as real-world ones. In step 3, we estimate the capacity of the deployment under study (real or synthetic) and compare it to the current traffic demand. In steps 4–6 we move to the time span going from now to 2020, and compare the capacity of present-day LTE networks to their expected load. Our objective is to assess whether and *which* enhancements, namely, MIMO and coordinated transmissions, will be necessary in order to meet the forecasted growth of traffic demand. The reason for choosing the above time horizon is twofold: first, reliable traffic forecasts are typically [6] limited to 2020. Furthermore, after 2020 5G is expected to bring major, even if still not completely clear, changes to network technologies. Notice however that the lessons we learn about cellular networks, the challenges they face, and the potential improvements thereto, represent valid insights for the design of 5G systems as well [7].

There are four main features that make our study unique:

- it relies on *generally available* traces (even if subject to a fee), unencumbered by non-disclosure agreements;
- the cell deployment we reconstruct from the traces at our disposal is cross-checked against independently-obtained data;
- its methodology is general, and applies to multiple network types and sources of information, as well as real-world and synthetic traces;
- it includes a way to generate synthetic traces;
- it accounts for the *evolution* of mobile networks and the traffic demand they serve, thus shedding some much-needed light on the strategies of mobile operators and the possible ways to improve them.

The rest of the paper is organized as follows. Sec. 2 discusses previous work, while Sec. 3 presents the traces that we use in our study, whose demand and deployment we characterize in Sec. 4. Sec. 5 presents our synthetic trace generation methodology, while Sec. 6 describes how we estimate the capacity of a given deployment, be it real or synthetic, and compare it to its current traffic demand. We then discuss the future development of both real and synthetic networks in Sec. 7. Our results, obtained for the real-world WeFi and OpenSignal traces, provide useful insights and guidelines for the design and enhancement of LTE networks. Finally, Sec. 8 draws our conclusions.

## 2 RELATED WORK

Our work is mainly related to studies on mobile network planning and enhancement, and to the body of work analysing real-world measurements of cellular network traffic.

Out of the vast literature existing on network planning, the works [8], [9] aim at optimizing the deployment of LTE base stations considering both coverage and capacity. A similar problem is addressed in, e.g., [10] in the case where multiple operators share the network infrastructure. The goal of our study is fundamentally different from all these works: we aim not to optimize infrastructure deployment or sharing, but to develop a methodology to characterise real-world LTE networks and study their performance and potential compared to the present and future traffic demand.

As far as capacity increase of mobile networks is concerned, several studies have focused on physical-layer techniques that can enable cellular networks to meet the growing traffic load. These include Coordinated Multipoint (CoMP), mmWave communications and MIMO [11]–[13]. Relevant to our work are also the studies on MIMO, such as [14]. We refer to the above works in order to get input values on the performance gains that these techniques are expected to yield.

Furthermore, several papers have appeared on the analysis of cellular traffic data traces, tackling different aspects. As an example, mobile traffic patterns of cellular towers are modeled through an empirical study in [15], while the geospatial and temporal dynamics of mobile traffic are studied in [16], [17]. User mobility and temporal activity patterns, as well as the usage of radio resources by different applications, are studied in [18], [19] for 3G networks. In [20], the aggregate temporal behavior of calling activity in a mobile phone network is used to infer daily mobility patterns in an urban area. Within this body of work, the spirit of [21] is the closest to that of our study. Indeed, [21] characterizes the operational performance of a 1-tier cellular network during high-profile crowded events, the experienced performance degradation in user service, and possible remedies.

Finally, our study is related to works that aim at obtaining stochastic models of networks and network dynamics. Among the most recent ones, [22] fits log-normal distributions to both the deployment (in space) and the demand (in time) of a mobile network in China. In a similar spirit, the authors of [23] study which, among several distributions, best matches the space patterns of real-world cellular networks in Italy. Both [22] and [23] employ different models for different types of areas, e.g., rural and urban ones; furthermore, [22] also separately models on- and off-peak hours. Both works obtain a very good correspondence between real-world and generated traces. With respect to them, we seek to improve the match in such local effects as the deployment changes between neighboring geographical areas, or the demand evolution between consecutive time periods.

A preliminary version of this study was included in our conference paper [24]. With respect to the conference version, this work has a wider scope, including the *generation* of cellular traces along with the usage thereof. Additionally, our performance evaluation includes a more detailed analysis of the root causes behind cellular network performance, and uses a wider set of real-world measurements.

## 3 INPUT DATA

We begin our analysis from two crowdsourced mobile network traces, coming from the WeFi [3] and OpenSignal apps [4], respectively. In particular, we consider the traces related to the San Francisco Bay Area, as depicted in Fig. 3. For the sake of simplicity, in the first part of our analysis we restrict our attention to the city and county of San Francisco, an $11 \times 11$ km$^2$ area marked with a black rectangle in Fig. 3.

This is a dense urban environment, challenging for any MNO to adequately serve. We focus on three major, US-wide MNOs, hereinafter randomly labeled from 1 to 3. Tab. 1

TABLE 1
The WeFi and OpenSignal San Francisco datasets

| Metric | WeFi | OpenSignal |
|---|---|---|
| Time of collection | March 2016 | |
| Number of records | 9 millions | 2 millions |
| Unique users | 7,182 | n/a |
| Unique Wi-Fi BSSIDs | 21,196 | 5,890 |
| Mobile operators (number of cells) | MNO 1 (7,998) MNO 2 (6,154) MNO 3 (2,338) | MNO 1 (2,123) MNO 2 (2,104) MNO 3 (1,294) |

summarizes the main features of both WeFi and OpenSignal (San Francisco only).

**WeFi.** WeFi collects information about the user's position, connectivity and activity. Each record within the dataset contains the following information:

- day, hour (a coarse-grained timestamp);
- anonymized user identifier and GPS position;
- MNO, cell ID[1], cell technology (e.g., 3G/4G) and local area (LAC) the user is connected to (if any);
- Wi-Fi network (SSID) and access point (BSSID) the user is connected to (if any) – but no indication on the Wi-Fi technology used;
- active app and amount of downloaded/uploaded data.

If the position of the user or the networks she is connected to change within a one-hour period, multiple records are generated. Similarly, one record is generated for each app that is active during the same period. The fact that records in the WeFi trace include mobility information allows us to track how much each user moves over time, and therefore to categorize users as static, pedestrian, or vehicular.

**OpenSignal.** The objective of OpenSignal is to construct a publicly available, operator-independent map of world-wide connectivity. To this end, users of the OpenSignal app volunteer to share their position and connectivity information, both cellular and Wi-Fi. Furthermore, users can decide to run speed tests, whose outcome – upload/download speed and latency – further enriches the map.

---

1. Cell IDs uniquely identify each cell within the MNO's network. They are not to be confused with physical cell IDs, i.e., integer numbers in $[0, 503]$ used for data scrambling on control channels.

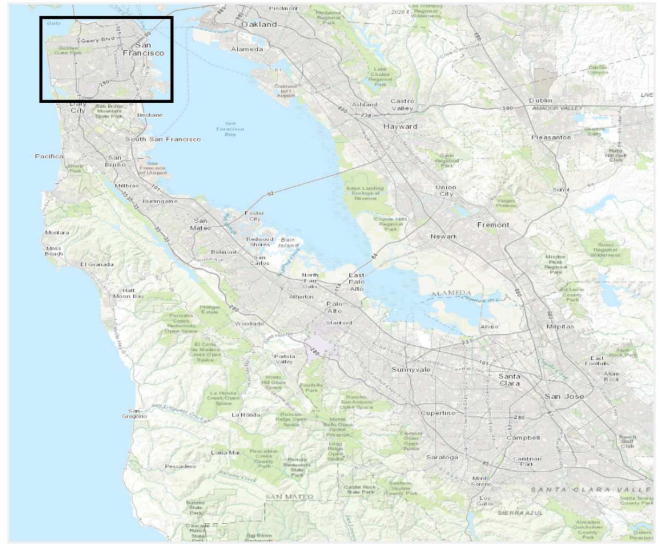

Fig. 3. The area covered by the WeFi and OpenSignal traces. The black rectangle corresponds to the city and county of San Francisco.

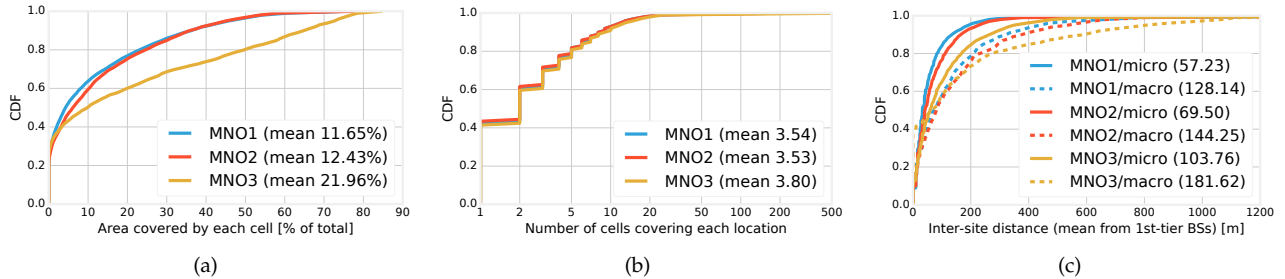

Fig. 4. Distribution of the area covered by each cell (a), of the number of cells covering each location (b), and of the inter-site distance for macro and micro-cells (c). In (a), covering 10% of the total area means a cell radius of roughly 2 km, 20% 2.8 km and 50% 4.5 km. Legends also report mean values within parentheses.

As highlighted in Tab. 1, OpenSignal data includes neither user identifiers nor application information. Furthermore, due to its smaller user base, it does not report some cells and Wi-Fi access points that are reported by WeFi. Notice that instead WeFi reports over 97% of the cells reported by OpenSignal, coming very close to being a superset thereto. Thus, we mostly base our study on the WeFi trace, owing to the larger amount of information it contains. We will use the OpenSignal trace whenever appropriate to complement and cross-check our results and observations.

**Availability and reproducibility.** Compared to traditional traces collected by MNOs, the datasets we use enable a more comprehensive vision of mobile networks, spanning different technologies (Wi-Fi and cellular) and multiple mobile networks. Another, non-technical, advantage is that our datasets are collected by commercial companies and are available under commercial terms. This makes our work easier to reproduce, and our findings easier to generalize. All the code needed to generate the results presented in this paper is available online at [25].

## 4 A data-driven look at LTE networks

Our purpose here is to use the information at our disposal to study the deployment of present-day LTE networks. To this end, we perform the steps 1–2A in Fig. 1. Note that, here and in the following, we focus on downlink data transfers, which are the most critical component of today's traffic – indeed, downloads account for over 81% of all traffic reported in the WeFi trace –, and are deemed to be so also in the future [6].

### 4.1 Network deployment

Let us first consider the number of cells that appear in the WeFi trace for each MNO, as per Tab. 1. We note that such a number is fairly high considering the geographical extension covered by the trace. We then look at the size of the cells, expressed as the fraction of the total area they cover.

In order to determine a cell coverage area, we rely on the locations from which users report being covered by the cell itself (i.e., they report the corresponding Cell ID). Specifically, every time a user reports being served by a certain cell at a certain time at a given location, we create a new *network access sample*. The cell coverage area is then computed as the *convex hull* of all network access samples corresponding to that Cell ID.

The results are presented in Fig. 4(a), which shows a quite high number of large cells. More than 50% of all cells cover over 10% of the whole area under study, and the coverage of the 10% biggest cells reaches (for MNO 1 and MNO 2) or exceeds (for MNO 3) half of the whole area. Recall that we are looking at $11 \times 11$ km$^2$, so a cell covering half of this region has a radius of 6.5 km – fairly commonplace for LTE macro-cells, even in urban scenarios. Also, a 10% coverage translates into roughly 2 km cell radius, thus MNOs have between 50% and 60% small/medium sized cells in their networks.

Since there are so many cells (see the last row of Tab. 1) and they are fairly large, the resulting coverage is very dense. As it can be seen from Fig. 4(b), 20% of all locations[2] are covered by more than 5 cells, and it is not uncommon to find areas covered by as many as 10 cells. Importantly, similar observations hold for both the WeFi and the OpenSignal trace.

To distinguish between macro- and micro-cells, we take the widely accepted [26], [27] value of 2 km as a watershed[3] between macro and micro-cells: cells whose range exceeds 2 km are classified as macro-cells, while the others (about 50-60%) are micro-cells. As we will see in Sec. 4.2, this choice is also backed by the comparison against third-party real-world data.

We then compute the inter-site distance for macro- and micro-base stations (BSs), expressed as the average distance of a macro (resp. micro) BS from its first-tier neighboring macro (resp. micro) BSs. The results are depicted in Fig. 4(c). It is interesting to compare the figures reported there with the inter-site distances of 5G use cases [28]. For micro-cells, our results are in agreement with the 50 m reported in [Tab. 8] [28]; for macro-cells, the inter-site distances we observe are shorter than the 200 m of [28, Tab. 7]. This is an interesting fact: densification is commonly thought of as a future trend, that will come to maturity as small cells (including femto-cells) become commonplace. Also, it is usually foreseen in two-tier scenarios, such as those recommended by 3GPP [29] for performance evaluation of cellular networks. Our data suggests that densification is already happening, at least in urban areas, and is carried out with tangled, medium to large-sized cells. This implies that not all results obtained in simplified reference scenarios may still hold in real-world networks: checking this is one of the goals of our study.

---

2. In order to show location-based results, the geographical area has been discretized by superimposing a 10-m grid thereto.

3. We remark that 2 km is the value widely considered (see, e.g., [26], [27]) as *maximum* radius of micro-cells.

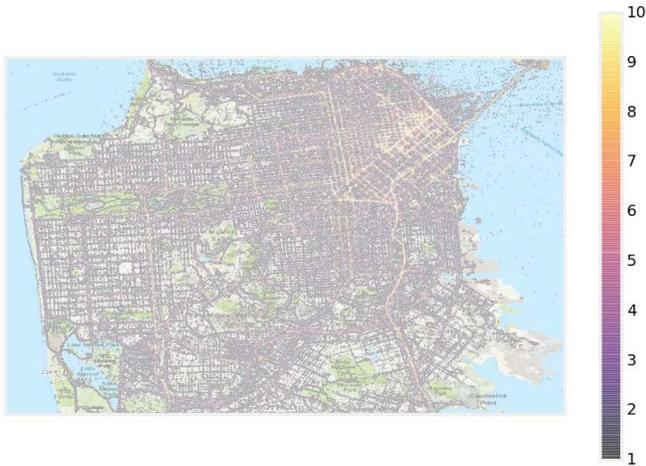

Fig. 5. Number of cells covering each location for MNO 2; lighter areas correspond to denser coverage. Maps for other operators (omitted for brevity) show a similar behavior.

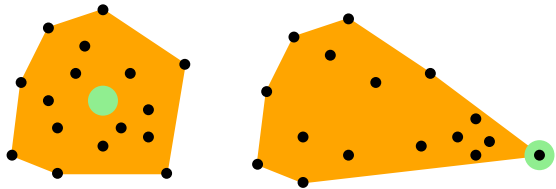

Fig. 6. Estimating the BS position. Black dots correspond to network access samples; orange polygons are the convex hull of the corresponding locations. Green dots represent the location we assign to the base station, which depends on the roundness of the convex hull. The polygon on the left has a roundness value of 0.68, hence we place an omnidirectional BS in the barycenter of the polygon (this is the case of some micro-cells). The polygon on the right has a roundness of 0.36, and we place a BS at one of its corners, chosen to be as close as possible to the positions of served users (this is the case of some micro-cells and all macro-cells).

Fig. 5 shows which zones exhibit a denser deployment: as expected, they turn out to be downtown areas (e.g., the financial district in the north-east) and the main thoroughfares (e.g., Market street immediately south of the financial district).

Next, we need to determine the location and type of the BSs serving each cell (step 2 in Fig. 1), a piece of information that is not present in our datasets. With regard to the BS location, it is important to note that real-world macro-cells typically employ tri-sectoral sites – a fact that is captured in the system models recommended by ITU [30], the CellMapper data [5] and confirmed by the shape of most of our macro-cells. Therefore, in the case of macro-cells, we place groups of three BSs at one corner of the coverage areas of as many cells, specifically, the one that minimizes the average distance from network access samples.

Microcells, on the other hand, are known to employ both directional and omnidirectional antennas. For each cell, we compute a *roundness* metric, defined as [31] $4\pi \frac{A}{P^2}$, where $A$ is the size of the cell coverage area, and $P$ is the perimeter thereof. Both $P$ and $A$ are computed directly from the coordinates of the vertices of the convex hull, through well-known formulas [32], as implemented by the `qhull` library [33]. The metric takes value $1$ for circles, and $0$ for segments. We then consider an omnidirectional antenna at the center of the coverage area for the cells with a roundness exceeding $0.5$, and a sector antenna for the others. Fig. 6 exemplifies how we construct the convex hulls and place the BSs for different types of cell. As discussed next, the above procedure allows a reliable cell classification, including cell type and sectorization.

### 4.2 Verifying the BS placement and type

In the following, we seek to assess how the BS placement and type that we obtain match the real ones, using the valuable data collected by the CellMapper project [5]. CellMapper is a crowdsourced effort aimed at building a cellular network map. It is similar in some aspects to OpenSignal, but it has a stronger emphasis on infrastructure, i.e., the type and location of BSs, rather than network coverage and speed. Also, unlike OpenSignal, they do not share their data with third parties, not even for a fee.

For our comparison, we randomly select[4] 100 Cell IDs from our real-world trace, and check:

- whether a cell with the same ID exists in CellMapper;
- how far away the position of its BS is from the one we consider;
- whether the number of sectors reported by CellMapper matches the one we establish;
- whether the frequency reported by CellMapper agrees with our classification (macro- or micro-cell).

Concerning the last item, recall that we classify cells as macro- or micro-cells based on the shape and size of their coverage area. CellMapper also collects information about the frequency used by each cell, so we can check whether the cells we classify as macro-cells use lower frequencies (700 to 900 MHz) and those we classify as micro-cells use higher ones (exceeding 1 GHz).

Tab. 2 summarizes the results of our comparison. The second column reports the fraction of Cell IDs from the the traces we use that exist in CellMapper. Reasons why that might not be the case include (i) cells that were never visited by CellMapper users; (ii) newly-built or decommissioned cells; (iii) cells whose ID was changed by the mobile operator for their own internal reasons – this is probably the case for MNO 3.

For the common cells, we can observe a remarkably good match between the BS location we obtain and that reported by CellMapper, as shown in the third column of Tab. 2, in terms of average distance between our estimated BS location and that reported by CellMapper. The fourth and fifth column show that the shape and size of coverage areas are very good indicators of the cell type and sectorization used by operators.

---

4. Since CellMapper does not provide access to its data, the comparison between Cell IDs has to be done manually, hence the comparison is limited to 100 cells.

TABLE 2
Comparison with CellMapper

| Operator | IDs match | Spatial error [distance] | Type match | Sectors match |
|---|---|---|---|---|
| MNO 1 | 63% | 32 m (macro) 5 m (micro) | 84% | 72% |
| MNO 2 | 72% | 21 m (macro) 12 m (micro) | 77% | 68% |
| MNO 3 | 39% | 38 m (macro) 8 m (micro) | 91% | 88% |

TABLE 3
Notation

| Symbol | Description |
|---|---|
| $a \in \mathcal{A}$ | Area types, e.g., urban or rural |
| $A(b) \in \mathcal{A}$, $A(t) \in \mathcal{A}$ | Area type of BSs $b \in \mathcal{B}$ and tile $t \in \mathcal{T}$ |
| $b \in \mathcal{B}$ | Base stations |
| $B(t)$ | Number of BSs deployed in tile $t \in \mathcal{T}$ |
| $k \in \mathcal{K}$ | Time slots, each corresponding to a one-hour period |
| $\mathcal{N}(t) \subseteq \mathcal{T}$ | Set of neighbors of tile $t \in \mathcal{T}$ |
| $t \in \mathcal{T}$ | Tiles, each corresponding to a $50 \times 50$ m$^2$ area |
| $T(b) \in \mathcal{T}$ | Location of BS $b \in \mathcal{B}$ |
| $\rho(b, k)$ | Real traffic demand at BS $b \in \mathcal{B}$ at time $k \in \mathcal{K}$ |
| $\delta(b, k)$ | Normalized traffic demand at BS $b \in \mathcal{B}$ at time $k \in \mathcal{K}$ |
| $\widehat{\delta}(b, k)$ | Synthetic, normalized traffic demand at BS $b \in \mathcal{B}$ at time $k \in \mathcal{K}$ |

We can conclude that the approach described in Sec. 4.1 is remarkably effective at reconstructing network deployment, cell types, and sectorization, using nothing but coverage information.

### 4.3 Summary

We performed the processing steps 1–2A in Fig. 1. Specifically, we characterized present-day LTE networks using WeFi and OpenSignal traces So doing, we observed a predominance of medium- to large-sized cells (Fig. 4(a)), resulting in a much denser deployment than we expected (Fig. 4(b)), with short inter-site distances (Fig. 4(c); the numbers in the legend therein represent the average values). We also observed how the deployment is especially dense in downtown areas and along the main thoroughfares (Fig. 5), and cross-checked the locations and types we obtain for cell BSs with those reported by the CellMapper project (Tab. 2).

## 5 GENERATING A NEW TRACE

In many cases, it would be desirable to generate a demand and/or mobility trace that exhibits the same features as those of real-world datasets, e.g., the ones we use, but that (i) relate to different areas and/or (ii) represent different realizations of the same traffic demand process. The latter case is especially relevant for performance evaluation, which ought to be carried out considering multiple, homogeneous but different, demand instances. Two situations are possible:

- the deployment is known (e.g., because public authorities mandate operators to disclose the locations of their BSs), but the demand is not;
- neither the demand nor the deployment is known.

We explain how we deal with these two cases in Sec. 5.1 and Sec. 5.2, respectively. After characterizing the computational complexity of our methodology in Sec. 5.3, we evaluate it against the real-world WeFi trace in Sec. 5.4.

Our notation is summarized in Tab. 3. We divide the space in discrete *tiles* $t \in \mathcal{T}$, each corresponding to a $50 \times 50$ m$^2$-square. We also divide the time into *slots* $k \in \mathcal{K}$, each corresponding to a one-hour period. For each tile, we know the *area type* $A(t) \in \mathcal{A}$, e.g., urban or rural; this information can be obtained from such sources as public census databases.

Base stations are represented by elements of a set, $b \in \mathcal{B}$; the tile a BS is located in is indicated as $T(b) \in \mathcal{T}$. We also define the type of the area where a BS $b$ is located (e.g., urban, suburban, rural) as $A(b) = A(T(b))$. Note that, in principle, we would need to consider also different types of BSs, e.g., serving micro/macrocells. However, for simplicity, here we present the procedure to generate a synthetic trace without distinguishing between BSs of different types; deployments including BSs of different types can be generated by simply repeating the procedure. Finally, we denote with $\rho(b, k)$ the real-world downlink data demand by users associated with BS $b$, during time slot $k$.

### 5.1 A synthetic demand trace

In many cases, mobile operators are required to disclose the location of their BSs, along with additional information such as power levels and used frequencies – often due to public concerns about "electromagnetic pollution".

From our viewpoint, this means that the BSs $b \in \mathcal{B}$ and their location $T(b)$ are known, and we have to determine their demand. Intuitively, this means devising a stochastic process whose realizations resemble the original demand contained in our traces.

A popular and successful approach to this problem is proposed in [22], [23]: to fit a distribution to the existing data (in our case, the data demand) and then extract samples thereof to represent the data demand at different time slots. However, the extracted samples would by necessity be i.i.d.; therefore, the synthetic demand would reproduce the *global* features of the actual one (average, variance), but not the *local* correlations between the demand at subsequent time slots.

In order to mitigate this problem, we opt for a discrete-time Markovian demand model, where the state is given by:

$$s = (A(b), d, h, \delta),$$

where:

- $A(b)$ is the type of area in which BS $b$ lies;
- $d$ is the day of week (Monday to Sunday; referred to as **DoW** in the pseudocode);
- $h$ is the hour (0 to 23);

- $\delta$ is the normalized traffic (expressed as an integer between 0 and 100).

Normalizing traffic values allows us to compare BSs that, in spite of having different levels of traffic, exhibit similar behaviors. Furthermore, using discrete traffic values makes it possible to have a finite set of states.

Alg. 1 shows how we train our Markovian model by estimating the state transition probabilities. In Line 3 and Line 4, leveraging the real-world data demand $\rho$, we compute the normalized traffic demands $\delta_i$ and $\delta_j$, referring to the current and next state. In Line 5 and Line 6 we compute the states themselves, each a 4-uple containing the type of area, day of week, hour, and normalized traffic demand. Line 7 updates a counter, $\mathtt{ctr}(s_i, s_j)$, keeping track of how many times we observe a transition from state $s_i$ to state $s_j$. These counters are then used in Line 9 to estimate the transition probabilities.

Notice how in Line 5 and Line 6 we use the functions **DoW** and **Hour** to compute the day of week and hour a certain time slot refers to. In general, there will be multiple time slots referring to the same weekday and hour, e.g., March 23rd at 3pm and March 30th at 3 pm. This allows us to observe more transitions from each state, and thus compute the transitions probabilities in a more reliable way.

Also notice that having different probabilities for different values of $A(b)$ is equivalent to building separate Markov chains for different area types – which is necessary, as traffic dynamics change significantly across urban, suburban and rural areas.

Given the transition probabilities, generating the traffic of a base station $b$ over time simply consists in generating an instance path over the relevant Markov chain, i.e., the one with the corresponding area type $A(b)$. As summarized in Alg. 2, we do so by first generating the current state $s_i$ (Line 3) and selecting the next state $s_j$ according to the appropriate probabilities (Line 4). Finally, we select the fourth element of the $s_j$ 4-uple as the synthetic traffic demand, denoted by $\widehat{\delta}$ (Line 5).

### 5.2 A synthetic deployment

In some cases, deployment information is not available, and we are only given the area under study (i.e., the tiles $t \in \mathcal{T}$ and the area types $A(t)$). Using this information, we have to

---

**Algorithm 1** Estimating the transition probabilities for the Markovian demand model
**Require:** $\mathcal{B}, \mathcal{K}, A(t), \rho$
1: **for all** $b \in \mathcal{B}$ **do**
2:   **for all** $k \in \mathcal{K}\colon k+1 \in \mathcal{K}$ **do**
3:     $\delta_i \leftarrow \lfloor 100 \frac{\rho(b,k)}{\max_{h \in \mathcal{K}} \rho(b,h)} \rfloor$
4:     $\delta_j \leftarrow \lfloor 100 \frac{\rho(b,k+1)}{\max_{h \in \mathcal{K}} \rho(b,h)} \rfloor$
5:     $s_i \leftarrow (A(b), \mathbf{DoW}(k), \mathbf{Hour}(k), \delta_i(b,k))$
6:     $s_j \leftarrow (A(b), \mathbf{DoW}(k+1), \mathbf{Hour}(k+1), \delta_j(b,k+1))$
7:     $\mathtt{ctr}(s_i, s_j) \leftarrow \mathtt{ctr}(s_i, s_j) + 1$
8: **for all** $s_i, s_j$ **do**
9:   $p(s_i, s_j) \leftarrow \frac{\mathtt{ctr}(s_i, s_j)}{\sum_{s_l} \mathtt{ctr}(s_i, s_l)}$
  **return** $p(s_i, s_j)$

---

**Algorithm 2** Generating the traffic demand
**Require:** $\mathcal{B}, \mathcal{K}, A(t), p(s_i, s_j)$
1: **for all** $b \in \mathcal{B}$ **do**
2:   **for all** $k \in \mathcal{K}\colon k-1 \in \mathcal{K}$ **do**
3:     $s_i \leftarrow (A(b), \mathbf{DoW}(k-1), \mathbf{Hour}(k-1), \delta(b,k-1))$
4:     $s_j \leftarrow$ **choose from** $\{s_j\}_j$ **with probability** $p(s_i, s_j)$
5:     $\widehat{\delta}(b,k) \leftarrow s_j[4]$
  **return** $\widehat{\delta}(b,k), \forall b \in \mathcal{B}, k \in \mathcal{K}$

---

**Algorithm 3** Estimating the probabilities $p(b, a, \beta)$ for the Bayesian deployment model
**Require:** $\mathcal{T}, A(t), B(t)$
1: **for all** $t \in \mathcal{T}$ **do**
2:   $\beta \leftarrow \left\lceil \frac{1}{|\mathcal{N}(t)|} \sum_{v \in \mathcal{N}(t)} B(v) \right\rceil$
3:   $\mathtt{ctr}(B(t), A(t), \beta) \leftarrow \mathtt{ctr}(B(t), A(t), \beta) + 1$
4: **for all** $b, a, \beta$ **do**
5:   $p(b, a, \beta) \leftarrow \frac{\mathtt{ctr}(b,a,\beta)}{\sum_x \mathtt{ctr}(x,a,\beta)}$
  **return** $p(b, a, \beta)$

---

create the deployment, i.e., the number $B(t)$ of BSs in each tile. Such synthetic deployment can then be used to obtain the synthetic demand $\widehat{\delta}(b, k)$ as we have seen in Sec. 5.1.

The intuition is similar to that of Sec. 5.1: we need a space distribution that resembles the one we observe in the real-world traces. We can describe a deployment through the number $B(t)$ of BSs existing in each tile:

$$B(t) = |\{b \in \mathcal{B} \colon T(b) = t\}|.$$

Most existing works [22], [23] solve this problem by fitting a distribution to the observed $B(t)$ values, and then extracting samples from it. However, as discussed earlier, such samples will be i.i.d., i.e., the number of BSs in a tile would be independent of the number of BSs in neighboring tiles around it. Experience and intuition tell us the contrary: in rural areas, having a BS in a neighboring tile means that there will probably not be any in the current one; conversely, in urban areas we are more likely to observe dense deployments spanning multiple contiguous tiles.

We can account for these effects through a Bayesian model, where the number $B(t)$ of BSs deployed at a tile $t$ depends upon:

- the area type $A(t)$ of $t$ itself, and
- the average number of BSs deployed in tiles neighboring $t$.

Specifically, we indicate as

$$p(b, a, \beta)$$

the probability of having $b$ BSs in a tile with area type $a$, given that the average number of BSs in the neighboring tiles is $\beta$. We train the Bayesian model by estimating the probabilities $p(b, a, \beta)$, as summarized in Alg. 3.

In Line 2, we compute the average number $\beta$ of BSs deployed in the tiles around the current tile $t$, given the set of neighbors $\mathcal{N}(t)$. We then update, in Line 3, the counter $\mathtt{ctr}(B(t), A(t), \beta)$ keeping track of how many tiles of type $A(t)$, whose neighbors have on average $\beta$ BSs deployed therein, have $B(t)$ BSs themselves.

We generate our synthetic deployment by using the $p$-probabilities to decide the number $B(t)$ of BSs to deploy in each tile. However, we need to take additional care in computing the $\beta$ values: these values depend on the previously-made decisions $B(t)$, but the decisions for some of the neighboring tiles might not have been made yet. In Line 2 of Alg. 4 we therefore initialize all $B(t)$ values to the flag value $-1$, marking tiles for which no decision has been made yet. Then, as long as there are tiles for which we still need to make a decision (Line 3), we proceed as follows.

In Line 4 we select the next tile to decide upon: among the tiles with no decisions (i.e., $B(v) < 0$), we choose the one with the highest number of neighbors *with* decisions. Then, in Line 5, we compute the set $\widehat{\mathcal{N}}$ of neighbors of $t$ for which a decision has been made (i.e., their $B$-value is not $-1$). Using this set, we compute the $\beta$-value in Line 6. Then, in Line 7, we decide the number $B(t)$ of BSs to place in tile $t$ using the $p$-probabilities returned by Alg. 3.

Note that the $B(t)$ values for the first tiles we decide about in Alg. 4 could be affected by the low number of neighbors with decisions, i.e., the small size of set $\widehat{\mathcal{N}}$. This issue can be addressed by running the algorithm multiple times, replacing the initialization in Line 2 with the previously obtained decisions.

### 5.3 Computational complexity

Our algorithms exhibit a low (namely, polynomial) computational complexity.

Specifically, Alg. 1 has a computational complexity of $O(|\mathcal{B}|^2|\mathcal{K}|^2)$. This can be seen by inspection of the algorithm itself: the nested loops between Line 1 and Line 7 run for exactly $|\mathcal{B}|(|\mathcal{K}| - 1)$ iterations, and within each loop at most 2 new states are created. The number of states is therefore $O(|\mathcal{B}||\mathcal{K}|)$. The final loop, starting in Line 8, runs once for each *pair* of states, hence $O(|\mathcal{B}|^2|\mathcal{K}|^2)$ times, and it represents the dominant contribution to the overall computational complexity. Alg. 2 has a complexity of $O(|\mathcal{B}||\mathcal{K}|)$, given by the two nested loops running for a total of $|\mathcal{B}|(|\mathcal{K}| - 1)$ iterations.

The complexity of Alg. 3 is dominated by the loop at the beginning of the algorithm, running for exactly $|\mathcal{T}|$ iterations, and its complexity is thus $O(|\mathcal{T}|)$. Similarly, each of the two loops in Alg. 4 runs at most $|\mathcal{T}|$ times, hence the algorithm complexity is $O(|\mathcal{T}|)$.

---

**Algorithm 4** Generating a synthetic deployment

**Require:** $\mathcal{T}, A(t), p(b, a, \beta)$
1: **for all** $t \in \mathcal{T}$ **do**
2: $\quad B(t) \leftarrow -1$
3: **while** $\exists v \in \mathcal{T} \colon B(v) < 0$ **do**
4: $\quad t \leftarrow \arg\max_{v \in \mathcal{T} \colon B(v) < 0} |\{n \in \mathcal{N}(v) \colon B(n) \geq 0\}|$
5: $\quad \widehat{\mathcal{N}} \leftarrow \{v \in \mathcal{N}(t) \colon B(v) \geq 0\}$
6: $\quad \beta \leftarrow \left\lceil \frac{1}{|\widehat{\mathcal{N}}|} \sum_{v \in \widehat{\mathcal{N}}} B(v) \right\rceil$
7: $\quad B(t) \leftarrow$ **choose** $x$ **with probability** $p(x, a, \beta)$
$\quad$ **return** $B(t)$

---

### 5.4 Testing the synthetic trace generation accuracy

Here we evaluate our synthetic trace generation procedure by assessing how closely the synthetic trace resembles the real-world WeFi trace. In order to observe a more varied set of urban, suburban and rural locations, we consider the *full* Bay Area trace shown in Fig. 3, instead of just the city of San Francisco.

We evaluate our procedure through the *de-facto* standard approach of $k$-fold cross-validation [34], [35]. In general, $k$-fold cross validation requires:

1) splitting the dataset into $k$ non-overlapping parts ("folds");
2) training $k$ different models, each using one of the folds as the testing set and the other ones as the training set;
3) computing the error metrics (e.g., RMSE) for each of the models;
4) averaging the error metrics.

Our goal is trace generation instead of traditional machine learning; therefore, we need to adapt the $k$-fold cross-validation approach and:

1) generate a different synthetic trace for each fold;
2) combine the $k$ synthetic traces, e.g., rounding the average number of BSs foreseen in a particular tile by the $k$ traces;
3) compare the combined synthetic trace with the real one.

It is also important to define how the folds should be generated, i.e., how the real-world traces should be divided into folds. Both time- and space-based divisions would result in potentially heterogeneous folds, e.g., having a different mix of urban and rural areas or referring to different times of the day. We instead opt for a *user-based* split, where each user ID is assigned to a fold with equal probability. By doing so, we ensure that all folds reflect the time and space features of the original trace.

Based on industry best practices [35] and tests run on reference datasets [36], we set $k = 10$.

#### 5.4.1 Demand

We begin from the simpler scenario where the infrastructure deployment is known, and we need to generate a synthetic traffic. We feed to Alg. 1:

- the BS locations, determined as explained in Sec. 4.1;
- the traffic they serve;
- the classification (urban, rural, suburban) of the area they serve, obtained by reverse-geocoding the locations and then querying the US census database [37].

Once we obtain the trained $p$-values, we run Alg. 2 and obtain the synthetic demand values $\widehat{\delta}(b, k)$.

One straightforward way to express how similar the synthetic demand $\widehat{\delta}$ is to the original demand $\overline{\delta}$ is to compute a global (i.e., over all BSs), per-hour traffic profile, as done in [22], [23]. This is generally defined as:

$$\Delta(h) = \sum_{b \in \mathcal{B}} \sum_{k \in \mathcal{K} \colon \mathbf{Hour}(k) = h} \delta(b, k), \quad h = 0 \ldots 23.$$

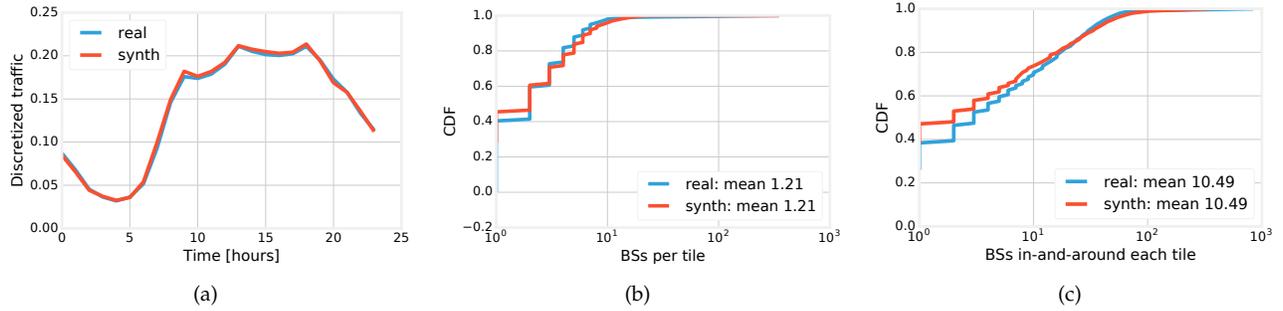

Fig. 7. Real and synthetic traces: traffic at different times of the day (a); distribution of the number of BSs in each tile (b) and number of BSs in the tiles surrounding each tile (c).

Then, we can compute the RMSE of the two traffic profiles:

$$RMSE_{\text{Hour}} = \sqrt{\frac{\sum_{h=0}^{23}(\bar{\Delta}(h) - \widehat{\Delta}(h))^2}{24}}. \quad (1)$$

As summarized in Tab. 4, the value for this metric is as low as $RMSE_{\text{Hour}} = 0.11$. Recalling that $\delta$-values are normalized traffic values between 0 and 100, this means that the gap between the total real and synthetic demand is as low as $0.11\%$. This is further confirmed in Fig. 7(a), showing an excellent agreement between real and synthetic demand.

The metric in (1) is, however, a little optimistic. By summing over all BSs and over all time slots corresponding to the same hour, we cannot check for inconsistencies in the amount of demand assigned to individual BSs. In other words, we do not know whether our synthetic trace confuses high- and low-traffic BSs. To check for this specific type of error, we (i) track the *total* (over time) amount of demand associated to each BS $\Delta(b) = \sum_{k \in \mathcal{K}} \delta(b,k)$ and (ii) compute the RMSE between $\Delta(b)$-values:

$$RMSE_{\text{B}} = \sqrt{\frac{\sum_{b \in \mathcal{B}}(\bar{\Delta}(b) - \widehat{\Delta}(b))^2}{|\mathcal{B}|}}. \quad (2)$$

As reported in Tab. 4, the $RMSE_{\text{B}}$ error is substantially lower than $1\%$. This shows that BSs that have a high traffic in the real trace are very likely to also have high traffic in the synthetic trace, and vice versa.

Finally, we compute the RMSE we obtain, by checking each BS at each time slot:

$$RMSE_{\text{BK}} = \sqrt{\frac{\sum_{b \in \mathcal{B}, k \in \mathcal{K}}(\bar{\delta}(b,k) - \widehat{\delta}(b,k))^2}{|\mathcal{B}||\mathcal{K}|}}. \quad (3)$$

TABLE 4
Demand errors

| Metric | Description | Value [%] |
|---|---|---|
| $RMSE_{\text{Hour}}$ (1) | Error margin between the real and synthetic hourly demand profiles (Fig. 7(a)) | 0.11 |
| $RMSE_{\text{B}}$ (2) | Error margin between the real and synthetic total demand assigned to each BS | 0.64 |
| $RMSE_{\text{BK}}$ (3) | Error margin between the real and synthetic demands for each BS and time slot | 7.94 |

The metric in (3) expresses how closely the two traces resemble each other, looking at individual time slots and base stations. The value we obtain for this very pessimistic metric is, as we can see from Tab. 4, $RMSE_{\text{BK}} = 7.94\%$. In other words, we can take any BS out of the thousands we have, any time slot out of hundreds, and our synthetic traffic will be, on average, only $8\%$ off from the real one.

We are able to obtain such a close resemblance between synthetic and real demand values thanks to our Markovian approach. By following a Markov chain, instead of extracting i.i.d. samples from a fitted distribution, we are able to account for the fact that (i) traffic demand evolves over time in a (more or less) smooth way, and (ii) individual BSs (even if they cover similar areas) serve different demands.

### 5.4.2 Deployment

We proceed in a similar way to evaluate the correspondence between the real deployment and the synthetic one. In particular, the input to Alg. 4 simply consists of the Bay Area topology, along with the area type information. The output is given by the number $\widehat{B}(t)$ of BSs assigned to each tile $t$.

Fig. 7(b) depicts the real and synthetic distribution of the number of BSs per tile. We can see that the average values almost exactly match, and the distributions exhibit a very good match.

We still need to ascertain whether our synthetic model is able to capture the correlation between deployments at *neighboring* tiles. To this end, in Fig. 7(c) we show the distribution of the number of BSs at each tile *and the neighboring ones*. Again, we can see almost the same average, and a very good match between the cumulative distribution functions (CDFs).

### 5.5 Summary

We presented a synthetic trace generation methodology. Our main intuition is to reproduce not only the global features of real-world traces, e.g., the total number of BSs, but also the local ones, e.g., the traffic differences between neighboring BSs. To this end, we use a Markovian models for the demand (Sec. 5.1) and a Bayesian model for the deployment (Sec. 5.2).

We evaluated the errors we make in the synthetic demand (Tab. 4, Fig. 7(a)) and the synthetic deployment (Fig. 7(b), Fig. 7(c)), and consistently found that we are able to reproduce global *and* local features of real-world traces.

TABLE 5
Frequencies assigned to MNOs. Source: FCC [38]

| Operator | 700 MHz | 1700 MHz | 1900 MHz |
|---|---|---|---|
| MNO 1 | 12 MHz, band 17 | — | 15 MHz, band 2 |
| MNO 2 | 5 MHz, band 12 | 15 MHz, band 4 | — |
| MNO 3 | 10 MHz, band 13 | 15 MHz, band 4 | — |

## 6 NETWORK CAPACITY AND TRAFFIC LOAD

Given the BS deployment and traffic demand, either from synthetic or real-world traces, we now perform step 3 of our procedure. Specifically, we aim at determining the capacity of the network and where it stands with respect to today's traffic load. We recall that, in order to answer these questions, we focus on downlink data transfers. We approach the above nontrivial task as follows:

1) computing the attenuation between geographical locations in the topology and BSs;
2) computing the signal-to-interference-plus-noise ratio (SINR) at each location in the trace, from every BS covering the location;
3) mapping the SINR onto the throughput associated with the LTE radio resource unit, i.e., a resource block (RB);
4) validating the results at the locations in the trace, against the traffic volume received by the users from their serving BS, and evaluating network overprovisioning.

**Signal propagation.** The first step is accomplished by exploiting the ITU models recommended for LTE networks serving urban areas [30]. In order to obtain an accurate model, we need to separately consider macro and micro-cells. Also, note that the models and the parameters we set are also in line with those foreseen for 5G urban environments [28].

- Micro-BSs, line-of-sight (LOS):
  $PL_{dB} = 40 \log d + 7.8 - 18 \log h_{BS} - 18 \log h_{UE} + 2 \log f$;
- Micro-BSs, non line-of-sight (NLOS):
  $PL_{dB} = 36.7 \log d + 22.7 + 26 \log f$;
- Macro-BSs NLOS:
  $PL_{dB} = 22 \log d + 28 + 20 \log f$.

In the equations above, $f$ is the frequency, $d$ is the distance between BS and user, $h_{BS}$ and $h_{UE}$ are the antenna heights of, respectively, BSs and users. We set $h_{BS} = 25$ m for macro-BSs, $h_{BS} = 10$ m for micro-BSs, and $h_{UE} = 1.5$ m [29], [30]. Following [30], we consider only the NLOS model for macro-BS and we use the LOS expression for micro-BS with probability

$$P_{LOS} = \min\{1, 18d^{-1}\}\left(1 - e^{-\frac{d}{36}}\right) + e^{-\frac{d}{36}}, \quad (4)$$

where $P_{LOS}$ is the probability that a given user is in line-of-sight of the serving BS.

Datasets typically do not include the frequencies used by BSs (parameter $f$ in the equations above). We look for this information in the FCC license database [38]. As an example, for the MNOs reported in the WeFi trace, we find that they all use frequencies at 700 MHz, and then some more at either 1700 or 1900 MHz (see Tab. 5).

These values naturally map into large and medium-sized cells, respectively: in the following, we consider that macro-cells use 700-MHz frequencies, while micro-cells use higher frequencies (whichever are available to their owner). We remark that the validity of this choice has been verified in Sec. 4.2. It is also worth stressing that FCC licenses can have a limited geographical scope, e.g., a single state or county. Tab. 5 only includes those licenses whose scope includes San Francisco; licenses valid for other areas are not included therein.

As aimed at by LTE MNOs, we initially assume frequency reuse factor of 1, i.e., all macro (micro, resp.) BSs use all the frequencies available to an MNO for macro (micro, resp.) cells. Also, in line with [29], [30] we assume a transmission power of 43 dBm for macro-cells, and 30 dBm for micro-cells. Using such a model, we can then compute the SINR that is experienced at each geographical location.

**Matching SINR with service data.** We now need to validate our signal propagation model by using the information included in the WeFi trace on the traffic volume served to the users. To this end, we can first map the SINR experienced by a user at a given location onto the amount of data that can be carried by one RB. We use experimental measurements [39] collected in the case of $2 \times 2$ MIMO – a fairly common setting in LTE networks –, and obtain the per-RB throughput. The number of available RBs is computed using the frequency allocation in Tab. 5. Then, in line with real-world LTE systems, we consider that proportional-fair scheduling is in place and obtain the throughput that can be offered at each location. Importantly, the experimental data in [39] shows that, in order to have a BS-user data communication, the SINR should be above -10 dB (which is also in accordance with the figures reported in [40]).

Fig. 8(a) depicts the distribution of the SINR for user-BS pairs that, in the WeFi trace, exchange data. The dashed lines therein refer to the case where we apply the path-loss equations to our data and set the frequency reuse factor to 1. We see that over 50% of communications that we observe in the WeFi trace are deemed impossible by our model, having SINR lower than -10 dB. This is a consequence of the dense deployment, which, under frequency reuse factor equal to 1, yields a very high interference. Note that decreasing the radius value taken as watershed between macro and micro-cells only worsen the situation (these results have been omitted for brevity). We therefore need to refine our model, in order for it to match the service coverage that emerges from the WeFi trace. Specifically, we need to account for interference mitigation techniques on the data plane, which in today's systems[5] mainly consist of flexible frequency reuse.

To this end, we relax the assumption on frequency reuse factor being equal to 1 and make *local*, per-BS decisions on which frequency bands to use. We adopt a hill-climbing approach, starting from those areas where users experience the lowest SINR. Then, given an area and initially setting the reuse factor to 1, we consider the individual BSs therein, starting with the ones with larger coverage. If we find it beneficial, we increase the reuse factor $K$ so that the BSs will use only a fraction $1/K$ of the available frequencies thus reducing the interference towards neighboring BSs and, at the same time, their own capacity. We found that, in order to match the downlink service data reported in the WeFi trace, $K$ should be typically increased to 3 for 8–18% of micro-cells and 40–48% of macro-cells, depending on the MNO. The high number of macro-cells and the fact that also micro-cells were involved, reflect the dense and tangled deployment we observe.

---

5. Note that the Almost-blank subframe (ABS) technique (i.e., one of the eICIC strategies) is not implemented yet in the networks of the considered MNOs. It will be considered as a way to enhance LTE networks in Sec. 7.

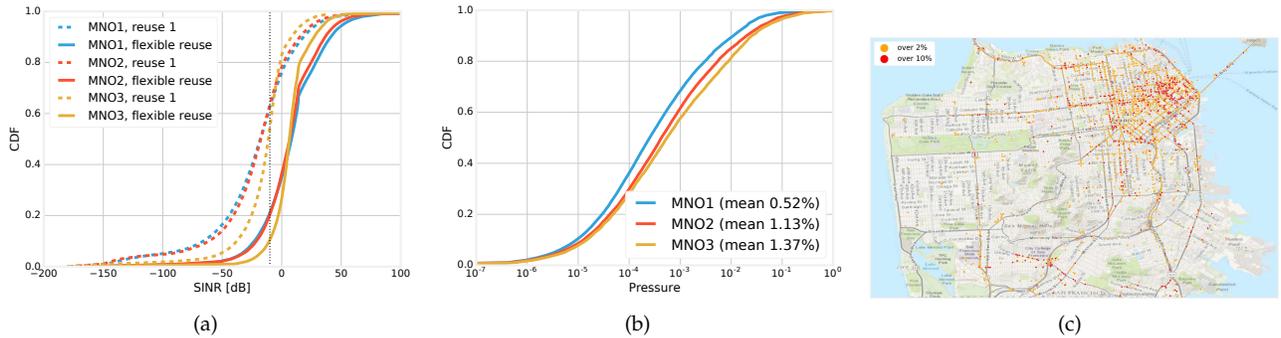

Fig. 8. Distribution of the SINR with (solid lines) and without (dashed lines) flexible frequency reuse (a); distribution of the pressure for different MNOs (b); locations where MNO 2 has a pressure exceeding 2% (yellow) or 10% (red) (c).

The final result is shown by the solid lines in Fig. 8(a), where the SINR at virtually every served location is above -10 dB. This means that the SINR that our model yields is in substantial agreement with the data transfers we observe from the trace. The fact that a few user-BS pairs still have a low SINR tells us that our model is slightly more conservative, a desirable property since, as detailed next, we are looking into worst-case, peak-time performance.

We now proceed and assess where the capacity of our networks stands with respect to their current traffic load. To do so, we need to find out the system *peak-time* load. A straightforward solution would be to consider the date and time with the highest total load, and use that snapshot as a reference. However, this would make us neglect that traffic load varies over both time and space. We thus construct a *combined* peak-load snapshot, where we consider the maximum load of each cell, and then combine together these local peak loads.

With reference to the San Francisco area, Fig. 8(b) depicts the distribution of the *pressure*, i.e., the ratio of the traffic demand to the throughput available at different locations. Consistently with the well-known fact that LTE networks are overprovisioned, pressure values average around 1%, and only exceptionally exceed 10%. Fig. 8(c) shows the moderate- and high-pressure areas for MNO 2 (maps for the other MNOs show similar results; they can be found in [25]). In accordance with common sense, we can clearly observe that downtown areas and main thoroughfares have higher pressure, and are thus more likely to become problematic in the future.

### 6.1 Summary

Our task in this section was to study a given deployment, be it real or synthetic, and ascertain to which extent the capacity thereof is suited to the traffic demand it currently has to serve. We leveraged real-world LTE facts, experimental measurements, ITU propagation models, and FCC license records, and we used real-world data as a validation tool. While simpler models would predict a very poor performance for such a dense deployment, we properly accounted for present-day interference mitigation techniques, obtaining SINR values (Fig. 8(a)) that are consistent with the data traffic reported in the WeFi trace. We also found that the network capacity far exceeds today's peak demand (Fig. 8(b)). As further confirmation of the correctness of our methodology, downtown areas and main thoroughfares are the locations where demand and capacity are the closest (Fig. 8(c)).

## 7 ENHANCING AND EXTENDING LTE NETWORKS

We now describe the processing steps corresponding to blocks 4–5 in Fig. 1, focused on the *future* demand and the ability of LTE networks to deal with it.

### 7.1 Future demand and pressure

Cisco is a prime source of information on future network demand. The figures below come from the 2016 edition of their Virtual Networking Index (VNI) report [6], which focuses on the 2016–2020 time period. The following are especially relevant to us:

- cellular traffic from non-mobile users will grow with a 57% compound annual rate (CAGR) [6, Fig. 2],
- cellular traffic from mobile users will grow with a 61% CAGR [6, Tab. 5],

which result in an overall annual rate of cellular traffic growth equal to 59%.

We then increase today's combined peak traffic, obtained in Sec. 6, using the CAGR figures provided by Cisco thus obtaining the projected future demand. Similarly, we multiply today's combined peak 3G demand by the same factors, and add that to the future LTE load. This way we account for the current trend of user traffic migrating from 3G to LTE[6]. Interestingly, as discussed in Sec. 3, the WeFi trace also contains reasonably accurate location information on individual users. Thus, in this case it is possible to distinguish between cellular traffic from mobile users and that from non-mobile users. Specifically, we can first mark as mobile any user that moves by more than 1 km in each one-hour period (needless to say, the same user can be marked as mobile in a time period and as non-mobile in others.) Then we can increase today's combined peak traffic by the CAGR values foreseen for non-mobile and mobile users.

---

6. As per the WeFi trace, data traffic is already much higher on LTE than on 3G.

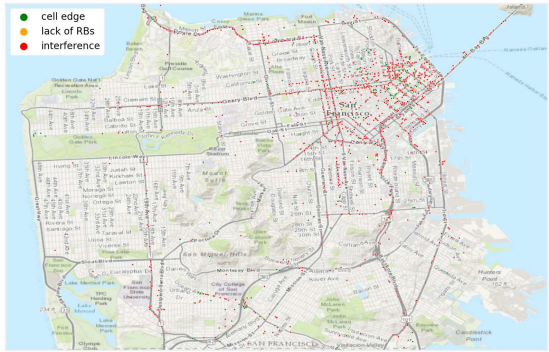

Fig. 9. Struggling locations for MNO 2 (WeFi trace) and the reason why they struggle.

We compare the demand values we obtain to the throughput that our networks can provide, and identify the *struggling* locations, i.e., the locations wherein the former exceeds the latter. The majority of locations will be able to deal with the future traffic – as one might expect from Fig. 8(b). However, as Fig. 9 exemplifies, each MNO will have several hundreds of struggling locations, where the network capacity cannot meet the demand, and action will be needed. We remark that the reason why certain locations are struggling is essentially twofold:

(1) the experienced SINR is low, hence each RB can carry a small number of bits and the provided data rate is not enough to support the future traffic demand. In particular, two factors contribute to a low SINR: (1.a) the location is highly interfered by neighboring BSs; (1.b) the signal received from the serving BS is weak, as in the case of cell-edge locations;

(2) the experienced SINR is satisfactory but the traffic load is exceedingly high, compared to the amount of available radio resources.

As an example, considering the WeFi trace, our model reveals that, quite consistently across the different MNOs, interference (case (1.a) above) is the main cause of struggle for more than 60% of locations, along with about 37% of locations being at the cell edge (case (1.b)). Struggling locations with a good SINR – higher than 5 dB – (case (2)) amount to only few percentage points. The different reasons why locations struggle in the case of MNO 2 are represented with different colors in Fig. 9, which also highlights that struggling locations include mainly downtown areas and thoroughfares. This is in agreement with the above percentages, as these areas exhibit a particularly dense network deployment (see Fig. 5) – hence many locations therein suffer high interference –, and they are burdened with high traffic demand. Thus, their SINR is insufficient to carry the required data load, as we can see from Fig. 8(c).

Below, we aim to "heal" struggling locations without extending the present-day network deployment.

### 7.2 Enhancing the network

In order to understand how the existing network can be improved to cope with the future traffic load, we investigate the following three strategies:

1) MIMO;
2) spectrum extension;
3) SINR increase.

We cascade the above strategies starting from those that aim to accommodate the additional traffic load without acting on the SINR (i.e., MIMO and spectrum extension). Then we target SINR increase and consider coordinated multipoint (CoMP) to mainly heal cell-edge locations, and almost-blank subframes (ABS) to mitigate interference. The reason for such an order is twofold. First, both traffic MIMO and spectrum extension are, at least partially, already in place, e.g., 2x2 MIMO in LTE. Thus, it is worth investigating to which extent they should be further pursued and enhanced. Second, CoMP and ABS increase the SINR at the expense of BS capacity; thus, they can be extensively applied when the network performance is not limited by the number of available RBs. This is confirmed by our experiments with the WeFi dataset, which gave the best results when the aforementioned order of actions was applied.

**MIMO.** Multiple-input-multiple-output (MIMO) is a technique that exploits multiple antennas at both the transmitter and the receiver in order to exploit multipath propagation. Present-day LTE networks already employ 2x2 MIMO, i.e., use two antennas at the transmitter and two at the receiver; in the following, we study the effects of updating to 8x8 MIMO. Choosing 8x8 MIMO is due do the fact that it is a comparatively mature technology, for which commercial equipment and real-world performance studies are available (e.g., [14]): it thus represents a first step mobile operators are taking towards next-generation networks, which are widely expected to feature massive MIMO.

We use experimental data from [14], reporting that moving from 2x2 MIMO to 8x8 MIMO results in a throughput improvement of 248%. Such figure is in agreement with other studies appeared in the literature. The third column of Tab. 6 and the second column of Tab. 7 summarize the effects of such a performance boost on the number of struggling locations: the results are encouraging, as more than 60% of all struggling locations, regardless the reason why they struggle, are healed.

**Spectrum extension.** MNOs are already extending the bandwidth available to LTE by *refarming* some of their spectrum: they are changing the destination of some frequency bands from GSM to LTE, and the same can be foreseen for 3G. We therefore focus on refarming as our spectrum extension strategy, and assess its efficacy.

Tab. 6 and Tab. 7 (fourth and third column, resp.) report how struggling locations fare, in the case of the WeFi trace, after 5 MHz (e.g., of GSM spectrum) are refarmed to LTE for each MNO, in addition to MIMO. Refarming 5 MHz yields a fairly high gain, especially for MNO 1. We then try to add an extra 5 MHz (e.g., from 3G spectrum) to LTE: doubling the new spectrum available to LTE yields substantially more healed locations. This suggests that spectrum refarming is a strategy worth pursuing aggressively, however - good news

TABLE 6
Number of struggling locations in the WeFi trace, healed by the strategies discussed in Sec. 7.2 when applied in the order reported below (percentages are given w.r.t. the number of locations struggling after the previous step). Green background highlights the strategies that heal over 40% of struggling locations, red background those that heal less than 20%. In the three rightmost columns, we consider that 5-MHz refarming is enabled

| Operator | Struggling | Healed by | | | | Residual |
|---|---|---|---|---|---|---|
| | | MIMO | Refarming | CoMP | ABS | |
| MNO 1 | 1083 | 770 (71%) | 140 (48%) [with 10 MHz: 145 (50%)] | 80 (46%) | 33 (35%) [with $K$=1: 36%] | 60 |
| MNO 2 | 2131 | 1329 (62%) | 203 (25%) [with 10 MHz: 210 (26%)] | 204 (34%) | 231 (58%) [with $K$=1: 58%] | 164 |
| MNO 3 | 1390 | 1007 (72%) | 131 (34%) [with 10 MHz: 140 (36%)] | 18 (7%) | 77 (33%) [with $K$=1: 34%] | 157 |

TABLE 7
How different strategies perform in healing locations struggling for different reasons (data from the WeFi trace; percentages given w.r.t. the number of locations struggling after the previous step)

| Reason | MIMO | Refarming 5/10 MHz | CoMP | ABS |
|---|---|---|---|---|
| Cell edge | 67% | 27/34% | 44% | 25% |
| Interference | 77% | 33/38% | 19% | 61% |
| Lack of RBs | 52% | 42/56% | 24% | 33% |

– 5–10 MHz are already enough to significantly improve the network performance. Tab. 7 also confirms the intuition that refarming mostly helps those locations that struggle due to lack of RBs, though around 30% of locations struggling for other reasons benefit from it as well.

At last we stress that, in spite of the above efforts, Tab. 6 reports a significant number of locations that are still struggling. These are the locations affected by very low SINR, compared to the forecasted traffic requirements. Refarming the spectrum means adding more RBs, but it does nothing to increase the amount of data each RB can carry – hence it may be not enough to heal locations with remarkably poor SINR. We also underline that such locations exhibit quite a high pressure already in the present (as per Fig. 8(c)), but the future increase in demand will exacerbate their situation. Consequently, below we proceed with two strategies that aim to increase the experienced SINR.

**Coordinated downlink transmissions.** Here we first[7] focus on CoMP [11], [41], which, using multiple BSs to serve a single location, helps to boost the power level of the useful signal and reduce interference at the same time. CoMP is therefore particularly beneficial to cell-edge locations, many of which appear to struggle.

We assign to each struggling location one additional BS, namely, the one from which the location receives the strongest signal, among those that (i) cover the location and (ii) have sufficient spare capacity. In the case of the WeFi trace, the results are reported in the fifth column of Tab. 6. For MNO 1 and MNO 2, CoMP heals about 40% of struggling locations. For MNO 3, instead, CoMP avails little, essentially because CoMP requires multiple BSs covering the same location, and this is less likely to happen for this MNO, as we can discern from Tab. 1. Furthermore, the fourth column in Tab. 7 confirms that CoMP is particularly beneficial to cell-edge locations.

Next, we consider ABS, a technique standardized by 3GPP but not currently implemented by the MNOs. According to ABS, BSs can refrain from transmitting in some subframes[8]. In our scenario, we make per-BS decisions on whether to implement ABS or not. If to be applied, in accordance with the surveyed literature [42], downlink transmissions are muted in 25% of subframes. We proceed in a simple hill-climbing fashion, starting from the BSs causing the most interference, skipping those lacking enough spare capacity, and stopping when implementing ABS stops being beneficial.

It is important to mention that, owing to the tangled deployment of our networks with no clear distinction of roles between macro and micro-cells, we considered that any BS can perform ABS if beneficial. However, from the WeFi trace we found that less than 5% of micro-cells need to perform ABS, versus 60–70% of macro-cells. This is in accordance with the fact that this technique is foreseen mainly for macro-cells, and it further validates the distinction we operate between micro and macro-cells.

The potential of ABS to improve performance is shown in the sixth column of Tab. 6, when applied on top of CoMP. ABS heals roughly 30% of the struggling locations for MNO 1 and MNO 3, and as many as 50% for MNO 2. Interestingly, although ABS was developed with classic two-tier deployments in mind, it works well also in the more tangled deployment we are observing. In addition, the last column of Tab. 7 shows that, although it is most effective for locations that struggle due to high interference, ABS heals several locations having too few available RBs and some cell-edge locations.

Finally, we check what happens if, while enabling ABS, we disable the flexible spectrum reuse we introduced in Sec. 6, i.e., we set $K=1$ for all BSs. In this case, ABS proves to be very effective: not only it makes up for the lack of flexible frequency reuse, but it also heals virtually the same number of struggling locations as before. This conforms with the notion that ABS and spectrum reuse serve mostly the same purpose in two different domains (time and frequency, respectively), and they are seldom both needed.

### 7.3 Summary

In this section, we proposed a methodology to evaluate how LTE networks can withstand their *future* load. Our first step was to construct a conservative, worst-case snapshot of such future load, using the Cisco projections [6]. As exemplified in Fig. 9 in the case of the WeFi trace, MNOs will be unable to provide the required capacity in more than one thousand locations each.

---

7. The order in which the techniques presented in this section are applied is the one yielding the best performance, although swapping them makes very little difference.

8. An LTE subframe is defined as a 1-ms time period.

We studied to which extent this situation can be eased by cascading MIMO, spectrum refarming, CoMP and ABS (Tab. 6), and we found that different strategies have different impact, also depending on the reason why locations struggle (Tab. 7). MIMO and refarming (especially, when an extra bandwidth of 10 MHz can be added) are both very effective on all locations. As we might expect, CoMP and ABS are mostly, although not exclusively, successful with cell-edge and highly-interfered locations, respectively.

## 8 CONCLUSIONS

By leveraging two large crowdsourced datasets (both available under commercial terms), we investigated the deployment of current, real-world LTE networks, and cross-checked them with independently-obtained data. Then, using ITU propagation models, FCC license records and experimental data, we have developed a methodology to assess the ability of LTE networks to support today's traffic load. Furthermore, we exploited projections on the growth of mobile data traffic and evaluated how LTE networks can cope with that.

Our study first unveils the quite dense and tangled deployment of macro and micro-cells of today's urban LTE networks, and provides a method to generate synthetic traces that closely resemble such deployment as well as real-world traffic demand. Importantly, such methodology is successfully applied to urban as well as suburban and rural environments.

## ACKNOWLEDGEMENT

We would like to thank Dr. Gian Michele Dell'Aera from Telecom Italia for the useful discussions. This work has received funding from the 5G-Crosshaul project (H2020-671598).

## REFERENCES

[1] "LTE-Advanced Pro: Part 3," July 2016.
[2] NGMN Alliance, "NGMN 5G white paper," https://www.ngmn.org/uploads/media/NGMN_5G_White_Paper_V1_0.pdf, accessed: 2016-07-11.
[3] "WeFi," http://www.wefi.com.
[4] "OpenSignal," http://www.opensignal.com.
[5] CellMapper. (2017) Cellular Coverage and Tower Map. https://www.cellmapper.net/.
[6] Cisco. Visual Networking Index 2016. http://www.cisco.com/c/en/us/solutions/collateral/service-provider/visual-networking-index-vni/mobile-white-paper-c11-520862.pdf.
[7] Qualcomm, "The 4G LTE network we have will get us the 5G network we need," 2016, https://www.qualcomm.com/news/spark/2016/02/24/4g-lte-network-we-have-will-get-us-5g-network-we-need.
[8] A. Guo and M. Haenggi, "Spatial stochastic models and metrics for the structure of base stations in cellular networks," *IEEE Transactions on Wireless Communications*, vol. 12, no. 11, pp. 5800–5812, Nov. 2013.
[9] H. Ghazzai, E. Yaacoub, M. S. Alouini, Z. Dawy, and A. Abu-Dayya, "Optimized lte cell planning with varying spatial and temporal user densities," *IEEE Transactions on Vehicular Technology*, vol. 65, no. 3, pp. 1575–1589, Mar. 2016.
[10] P. Di Francesco, F. Malandrino, T. K. Forde, and L. A. DaSilva, "A sharing- and competition-aware framework for cellular network evolution planning," *IEEE Transactions on Cognitive Communications and Networking*, vol. 1, no. 2, pp. 230–243, June 2015.
[11] B. Mondal, E. Visotsky, T. A. Thomas, X. Wang, and A. Ghosh, "Performance of downlink comp in lte under practical constraints," in *IEEE International Symposium on Personal, Indoor and Mobile Radio Communications (PIMRC)*, Sept. 2012, pp. 2049–2054.
[12] G. C. Alexandropoulos, P. Ferrand, J. m. Gorce, and C. B. Papadias, "Advanced coordinated beamforming for the downlink of future LTE cellular networks," *IEEE Communications Magazine*, vol. 54, no. 7, pp. 54–60, July 2016.
[13] C. Dehos, J. L. Gonzlez, A. D. Domenico, D. Ktnas, and L. Dussopt, "Millimeter-wave access and backhauling: the solution to the exponential data traffic increase in 5G mobile communications systems?" *IEEE Communications Magazine*, vol. 52, no. 9, pp. 88–95, Sept. 2014.
[14] K. Werner, H. Asplund, B. Halvarsson, A. K. Kathrein, N. Jalden, and D. V. Figueiredo, "LTE-A field measurements: 8x8 MIMO and carrier aggregation," in *IEEE VTC Spring*, 2013.
[15] H. Wang, F. Xu, Y. Li, P. Zhang, and D. Jin, "Understanding mobile traffic patterns of large scale cellular towers in urban environment," in *ACM Conference on Internet Measurement Conference (IMC)*, 2015, pp. 225–238.
[16] D. Willkomm, S. Machiraju, J. Bolot, and A. Wolisz, "Primary users in cellular networks: A large-scale measurement study," in *IEEE DySPAN Symposium on New Frontiers in Dynamic Spectrum Access Networks*, Oct. 2008, pp. 1–11.
[17] M. Z. Shafiq, L. Ji, A. X. Liu, J. Pang, and J. Wang, "Geospatial and temporal dynamics of application usage in cellular data networks," *IEEE Transactions on Mobile Computing*, vol. 14, no. 7, pp. 1369–1381, July 2015.
[18] U. Paul, A. P. Subramanian, M. M. Buddhikot, and S. R. Das, "Understanding traffic dynamics in cellular data networks," in *IEEE INFOCOM*, Apr. 2011, pp. 882–890.
[19] J. Yang, Y. Qiao, X. Zhang, H. He, F. Liu, and G. Cheng, "Characterizing user behavior in mobile internet," *IEEE Transactions on Emerging Topics in Computing*, vol. 3, no. 1, pp. 95–106, Mar. 2015.
[20] P. Dazzi, M. Dell'Amico, L. Gabrielli, A. Lulli, P. Michiardi, and M. Nanni, "Improving population estimation from mobile calls a clustering approach," in *IEEE International Symposium on Computer and Communications (ISCC)*, Messina, June 2016.
[21] M. Z. Shafiq, L. Ji, A. X. Liu, J. Pang, S. Venkataraman, and J. Wang, "Characterizing and optimizing cellular network performance during crowded events," *IEEE/ACM Transactions on Networking*, vol. 24, no. 3, pp. 1308–1321, June 2016.
[22] J. Ding, X. Liu, Y. Li, D. Wu, D. Jin, and S. Chen, "Measurement-driven Capability Modeling for Mobile Network in Large-scale Urban Environment," in *IEEE MASS*, 2016.
[23] L. Chiaraviglio, F. Cuomo, M. Maisto, A. Gigli, J. Lorincz, Y. Zhou, Z. Zhao, C. Qi, and H. Zhang, "What is the Best Spatial Distribution to Model Base Station Density? A Deep Dive into Two European Mobile Networks," *IEEE Access*, 2016.
[24] F. Malandrino, C.-F. Chiasserini, and S. Kirkpatrick, "Understanding the present and future of cellular networks through crowd-sourced traces," in *IEEE WoWMoM*, 2017.
[25] Code. https://dl.dropboxusercontent.com/s/ssqr1u6xxm4rpnq/index.html.
[26] *Small Cells - Technologies and Deployment, Second and Expanded Edition*. John Wiley & Sons, 2014.
[27] S. Peric and T. Callahan, "Challenges with Microcell Deployment & Configuration," 2013, https://www.wirelessdesignmag.com/article/2013/04/challenges-microcell-deployment-configuration.
[28] EU Project METIS-II: Mobile and wireless communications Enablers for Twenty-twenty (2020) Information Society-II, "Living document on 5G-PPP use cases and performance evaluation models," accessed: 2016-07-11. [Online]. Available: https://5g-ppp.eu/wp-content/uploads/2014/02/5G-PPP-use-cases-and-performance-evaluation-modeling_v1.0.pdf
[29] 3GPP Technical Report 36.814, "Further advancements for E-UTRA physical layer aspects," 2010.
[30] ITU. Recommendation ITU-R P.1411-1. https://www.itu.int/dms_pubrec/itu-r/rec/p/R-REC-P.1411-1-200102-S!!PDF-E.pdf.
[31] NISR. Roundness Measurements. http://www.itl.nist.gov/div898/handbook/mpc/section3/mpc344.htm.
[32] M. De Berg, M. Van Kreveld, M. Overmars, and O. C. Schwarzkopf, *Computational geometry*. Springer, 2000.
[33] C. B. Barber, D. P. Dobkin, and H. Huhdanpaa, "The Quickhull algorithm for convex hulls," *ACM Transactions on Mathematical Software*, 1996.

[34] J. Friedman, T. Hastie, and R. Tibshirani, *The elements of statistical learning*. Springer, 2001, vol. 1.
[35] I. H. Witten, E. Frank, M. A. Hall, and C. J. Pal, *Data Mining: Practical machine learning tools and techniques*. Morgan Kaufmann, 2016.
[36] D. Anguita, L. Ghelardoni, A. Ghio, L. Oneto, and S. Ridella, "The K in K-fold cross validation," in *Proceedings, European Symposium on Artificial Neural Networks, Computational Intelligence and Machine Learning. Bruges (Belgium)*, 2012, pp. 441–446.
[37] US Census. Urban Area Relationship Files. https://www.census.gov/geo/maps-data/data/ua_rel_layout.html.
[38] FCC. License View API. https://www.fcc.gov/general/license-view-api.
[39] FP7 IP EARTH, "Energy efficiency analysis of the reference systems, areas of improvements and target breakdown," 2010.
[40] H. Holma and A. Toskala, *WCDMA for UMTS: HSPA Evolution and LTE*. John Wiley & Sons, 2010.
[41] A. Marotta, K. Kondepu, F. Giannone, S. Doddikrinda, D. Cassioli, C. Antonelli, L. Valcarenghi, and P. Castoldi, "Performance evaluation of CoMP coordinated scheduling over different backhaul infrastructures: A real use case scenario," in *IEEE ICSEE*, 2016.
[42] S. Deb, P. Monogioudis, J. Miernik, and P. Seymour, "Algorithms for enhanced inter cell interference coordination (eICIC) in LTE HetNets," *IEEE/ACM Transactions on Networking*, 2014.

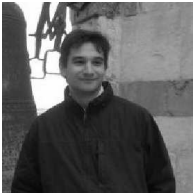

**Francesco Malandrino** earned his Ph.D. in 2012 from Politecnico di Torino, Italy, where he is currently a post-doc. Before his current appointment, he held short-term positions at Trinity College, Dublin, and at the Hebrew University of Jerusalem as a Fibonacci Fellow. His interests focus on wireless and vehicular networks and infrastructure management.

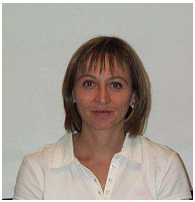

**Carla-Fabiana Chiasserini** (M'98, SM'09) graduated in Electrical Engineering (summa cum laude) from the University of Florence in 1996. She received her Ph.D. from Politecnico di Torino, Italy, in 2000. She has worked as a visiting researcher at UCSD in 1998–2003, and she is currently an Associate Professor with the Department of Electronic Engineering and Telecommunications at Politecnico di Torino. Her research interests include architectures, protocols, and performance analysis of wireless networks. Dr. Chiasserini has published over 230 papers in prestigious journals and leading international conferences, and she serves as Associated Editor of several journals.

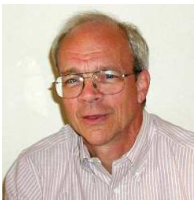

**Scott Kirkpatrick** (SM '80, F '91) has been a professor in the computer science department at the Hebrew University, in Jerusalem, Israel, since 2000. Before that he was at IBM Research, Yorktown Heights, where he did research in physics and engineering, developing simulated annealing and IBMs first tablet computers. Professor Kirkpatrick is a Fellow of the APS, the AAAS, the IEEE, and the ACM, has written over 100 papers and holds more than 10 patents. He holds an AB from Princeton University and PhD (in physics) from Harvard University.